\documentclass[11pt,a4paper]{article}

\usepackage{bm,amsmath,amssymb}
\usepackage{graphicx}
\usepackage{lineno,hyperref}
\usepackage{fancyhdr}

\newcommand{\kv}{$k$-vector}
\newcommand{\B}[1]{{\bm #1}}
\newcommand{\ds}{\displaystyle}
\newcommand{\dd}{\; \text{d}}

\fancypagestyle{firststyle}
{
	\fancyhf{}
	\fancyhead[RE,LO]{Computing in Science \& Engineering, Vol. 21, No. 1, pp. 94-107, 2019. \\
		doi: 10.1109/MCSE.2018.2882727}
	\fancyfoot[RE,LO]{doi: 10.1109/MCSE.2018.2882727}
	\fancyfoot[LE,RO]{Page \thepage}
}

\begin{document}
	
\title{Random Sampling using \kv}

\author{David Arnas\thanks{Assistant Professor, Centro Universitario de la Defensa Zaragoza, Crta. Huesca s/n, 50090 Zaragoza, Spain. Email: \textsc{darnas@unizar.es}}, \, Carl Leake\thanks{Graduate Student, Aerospace Engineering, Texas A\&M University, College Station, TX. E-mail: \textsc{leakec@tamu.edu}}, \, Daniele Mortari\thanks{Professor, Aerospace Engineering, Texas A\&M University, College Station, TX. E-mail: \textsc{mortari@tamu.edu}}}
	
\date{}	

\maketitle

\thispagestyle{firststyle}

\begin{abstract}
	This work introduces two new techniques for random number generation with any prescribed nonlinear distribution based on the \kv\ methodology. The first approach is based on inverse transform sampling using the optimal \kv\ to generate the samples by inverting the cumulative distribution. The second approach generates samples by performing random searches in a pre-generated large database previously built by massive inversion of the prescribed nonlinear distribution using the \kv. Both methods are shown suitable for massive generation of random samples. Examples are provided to clarify these methodologies.
\end{abstract}

\section{Introduction}\label{sec:intro}

Developments in the fields of statistics and probability have provided powerful tools that are useful for dealing with a large number of problems in science and engineering. Examples of application of these tools appear in computer simulation, particle physics, fluidynamics, cryptography, optimization techniques, biology, or finance. Mathematically, one of the most important concepts used in statistics and probability is the random variable. However, although it is a simple concept, its exact implementation in a computer has many difficulties. In fact, random number generation still exists as an open problem in computer science today, which has led to the development of a multitude of techniques that provide different solutions to this problem.

A general classification of random number generators can be made by investigating the source of the random numbers. Some random number generators are based on hardware. These techniques create genuinely random numbers, because the source of the numbers comes from physical magnitudes that change constantly. Much more common techniques create random numbers using algorithms (software). These algorithms are computer programs that generate pseudo-random sequences of numbers following a series of computer instructions. This means that, in practice, it is possible to reproduce the sequence of numbers generated if the starter or seed in known beforehand. However, these algorithms tend to be faster than hardware-based random number generators, and are also easier to implement, while they maintain a behavior that is hardly distinguishable from a true random sequence in the applications where they are used.

In general, random number generators produce samples distributed uniformly in a given range. This generation can lead to, for example, integer numbers between two given boundaries or to real numbers in the range $\left[0, 1\right]$. Nevertheless, there are many applications where generating samples following a given probability density function is required: examples include computer simulations, Monte Carlo simulations, and quantum physics. In those cases, it is necessary to transform the results from the uniform distribution, obtained from the random number generators, to a given probability density function. In that respect, several methodologies exist that deal with this problem, and the two most well known solutions are ``inverse transform sampling" and ``rejection sampling."

Inverse transform sampling is based on the idea of transforming a set of sample numbers from a given distribution into a different one. This is performed by the inversion of the cumulative probability density function of the objective function for all points generated in the original uniform distribution. This means that a function inversion must be done for each of the samples generated, which depending on the objective function, could be a costly computational process.

On the other hand, rejection sampling (also known as acceptance-rejection sampling) is based on generating both a random sample and an auxiliary random number that will be compared with the probability density function sought. Then, if the value of the probability density function is larger than the random number generated, the sample is accepted. Otherwise, the sample is rejected and an additional pair must be generated. This means that, in its most basic form and in order to generate a sample, the algorithm is required to produce two random numbers and evaluate the probability density function at least once.

In this work, we introduce two new methodologies that use the uniform distributed samples, created via a random number generator, to generate a set of samples following a given probability density function. Both methodologies are based on the \kv, an orthogonal range searching technique originally devised for the Star Identification (Star-ID) problem in spacecrafts \cite{Original}. The technique has been particularly successful in solving the Star-ID problem due to the minimal computational capability required by the on-board computer. For this reason, the \kv\ has been adopted by many different Star-ID algorithms, including ``Pyramid" \cite{pyramid}, the current state-of-the-art algorithm. The key idea of the \kv\ is to describe the non-linearity of a sorted database using a vector of integers containing the position of certain elements of the database, the \kv. The most important aspect of this range searching technique is that \emph{the searching complexity is independent from the database size}, making this technique particularly suitable for very large static databases.

The idea of using this technique to solve a variety of problems was initially proposed in Ref. \cite{Rogers}, including its possible applications in nonlinear function inversion, which is analyzed in detail in Ref. \cite{inversion}. The focus of this work is to use the \kv\ for random number generation. In this manuscript, we propose two different methodologies that deal with the problem of generating samples following a given probability density function. Both techniques require a one time preprocessing effort for each probability density function considered, where an auxiliary database containing the \kv\ has to be generated and stored. However, once this preprocessing is performed, the algorithm is able to generate the random samples using a very fast process. In fact, both algorithms are specially suited for the creation of a toolbox devised for random number generation.

The first technique is based on the inverse transform sampling property, where the \kv\ is used to invert the cumulative probability density function by performing a search in an auxiliary database containing points of the cumulative probability density function. This transforms the distribution in an extremely fast process. Additionally, this technique is further improved by the introduction of the \emph{optimal} \kv, a modification of the \kv\ methodology that always retrieves the same number of elements (e.g., 1 or 2) for a prescribed searching range, no matter where the search is performed. This means that one can, for example, always obtain two points per root to use in root solvers (e.g., bisection, regula-falsi) requiring root bracketing. Once the optimal \kv\ is built, and if the database is large enough, the methodology \emph{does not need to evaluate the function} and the process \emph{does not need to perform any additional searches}. Furthermore, this technique can be adapted to the size of the available memory and to the speed requirements of the application considered \cite{MAKV}.

The second proposed methodology is based on randomly searching indices in a pre-computed database that contains a distribution of points following the probability density function sought. Here, the \kv\ methodology is used to generate these points by a massive inversion of the distribution function. At each function inversion, the algorithm generates a number of samples proportional to the distance between the roots, which also generates the auxiliary database following the probability density function. Once this distribution is built, the methodology only requires generating a random index and retrieving the element associated with it from the auxiliary database.

\section{Preliminaries}\label{sec:preliminaries}

This section summarizes some concepts that are used in this work. First, a summary of the methodologies for inverse transform sampling and rejection sampling are presented, as they are the most common adopted techniques. Second, the general procedure of the \kv\ methodology is summarized.

\subsection{Inverse transform sampling}\label{subsec:inverse}

Inverse transform sampling can be summarized as follows \cite{devroye}: let $x$ be a random variable with continuous and invertible cumulative distribution function, $f$. Let $f^{-1}$ be its inverse function. Let $y$ be a random variable such that $y = f (x)$. If $y$ is uniformly distributed in $[0, 1]$, then $x = f^{-1}(y)$ follows the distribution in $f$.

In this approach a uniform distribution of points, $y \in [0,1]$, is generated and then the inverse of the cumulative distribution function must be computed to generate the samples, $x$ (following the density distribution $f$). This approach requires a large number of inversions of $f$ at some computational cost, and for this reason, this method may be computationally inefficient for many distributions and other methods are preferred.

An example of improving the performance of the basic method can be seen in Ahrens \cite{ahrens} where a guide table is set up before performing the sampling computation to speed up the inversions the method requires.

\subsection{Rejection sampling}\label{subsec:rejection}

Rejection sampling \cite{neumann}, also known as acceptance-rejection method, is a group of methodologies based on Monte Carlo methods that generates random distributions of samples following a given density function. The rejection sampling method is based on the following property \cite{devroye}: let $X$ be a random variable in $\mathbb{R}^n$ with density $f$. Let $U \in [0, 1]$ be a uniform random variable, and let $S = \left\{(x, u):x \in \mathbb{R}^n, \: 0\leq u\leq c \, f(x)\right\}$, where $c$ is an arbitrary constant. Then, $\left(X, c \, U f(X)\right)$ is uniformly distributed in $S$.

The rejection sampling method, in its basic form, requires defining a function $g (x)$, that bounds the initial distribution $f$, and a constant $c$ where:
\begin{equation}\label{rejection}
    f (x) < c \, g (x),
\end{equation}
for all $x$, where $g$ and $c$ are related to the target distribution $f$. More precisely, $g$ must have heavier tails and sharper infinite peaks than $f$ and fulfill Eq.~\eqref{rejection} for the interval of $x$ considered.

This methodology requires checking, for each sample, if the density function $f$ is above or below the actual value obtained in the computation, which leads to a number of samples that have to be rejected in the process. Thus, algorithms aim to increase the ratio of acceptance to rejection while simultaneously keeping the computation of $c \, g (x)$ as efficient as possible.

There are many algorithms based on this concept and that deal with the sampling problem with different approaches. Examples of them are Neumann \cite{neumann}, Ziggurat \cite{ziggurat}, Adaptive Rejection Metropolis Sampling \cite{arms}, and Slice Sampling \cite{slice}.

\subsection{Background on the \kv}\label{subsec:kvector}

The range searching problem is based on the retrieval, in a database of size $n$, of all the elements that are contained in a given interval $[y_a, \: y_b]$ where $y_a$ and $y_b$ are the lower and upper bounds of the range, respectively. Some of the most common searching algorithms are based on binary search, which has a complexity of $\mathcal{O}(2\log_2 n)$, and search by hashing, which is fast for the average case but can be linear for the worst case due to the collision problem affecting the method. The \kv\ range searching technique is as fast as a hashing method (with best case complexity of $\mathcal{O}(3)$),  but it does not present the collision problem. However, the \kv\ may return some of the elements that are closest to, but not included in the range. In addition, the \kv\ requires a very fast preprocessing effort: store the elements of the \kv\ by reading the sorted database sequentially. The \kv\ is particularly suitable for searching in large databases, and its complexity asymptotically approaches $\mathcal{O}(3)$ as more memory is available to store the \kv\ elements \cite{MAKV}.

The \kv\ is built using a sorted database ordered in the ascending mode. Let $n$ be the number of elements of the database, $\B{y}(i)$ be the $i$-th element, and $\B{s}$ the vector containing the sorted database $\B{y}$, that is,
\begin{equation}
    \B{s}(i) \leq \B{s}(i+1), \quad \forall i\in[1,n-1].
\end{equation}
where $y_{\min} = \B{s} (1)$ and $y_{\max} = \B{s} (n)$. Then, $\B{I}$ can be defined as the sorting indexes vector relating $\B{y}$ and $\B{s}$, that is,
\begin{equation}
    \B{s}(i) = \B{y}(\B{I}(i)) \quad \text{where} \quad i = \{1, 2, \cdots, n\}.
\end{equation}

The \kv\ is a vector of indexes containing information about the nonlinearity of the sorted database, in other words, its variation with respect to a line. This line is the line connecting the minimum and maximum values of the sorted database ($y_{\min}$ and $y_{\max}$). However, in order to include all the elements given the rounding and machine errors, the line is slightly extended to connect the points $[1, y_{\min} - \delta\varepsilon]$, and $[n, y_{\max} + \delta\varepsilon]$, where,
\begin{equation}
    \delta\varepsilon = (n - 1) \varepsilon,
\end{equation}
and $\varepsilon$ is the relative machine precision ($2.22\times 10^{-16}$ for double precision arithmetic). Therefore, the \kv\ line equation is,
\begin{equation}\label{kline}
    \begin{split}
    \B{y}_l(i) =& \; m \, (i - 1) + q = \\ ~ =& \, \dfrac{y_{\max} - y_{\min} + 2\delta\varepsilon}{n - 1} (i - 1) + y_{\min} - \delta\varepsilon,
    \end{split}
\end{equation}
where,
\begin{equation}
    m = \dfrac{y_{\max} - y_{\min} + 2\delta\varepsilon}{n - 1}, \quad q = y_{\min} - \delta\varepsilon,
\end{equation}
and $i = \{1, 2, \cdots, n\}$. The line defined in Eq. (\ref{kline}) has the purpose to set a series of reference levels, $\B{y}_l(i)$, that are used to generate the \kv.

The \kv\ ($\B{k}$) stores the nonlinearity of the sorted database by counting the number of elements that are below a given level defined by Eq. (\ref{kline}). This is equivalent to $\B{k}(i) = j$, where $j$ is the greatest index that fulfills $\B{s}(j) \leq \B{y}_l(i)$, that is,
\begin{equation}
    \B{k}(i) = \max\left(\{j \mid \B{s}(j) \leq \B{y}_l(i)\}\right).
\end{equation}
This also implies that $\B{k}(1) = 0$ and $\B{k}(n) = n$, as there are no elements below $\B{y}_l(1)$ nor above $\B{y}_l(n)$.

Once the \kv\ is built, the information that it contains as well as the line defined in Eq. (\ref{kline}) are used to find the elements of the database that are inside the range $[y_a, \, y_b]$. Let $k_a$ and $k_b$ be the two indexes that correspond to the boundary of the database $[y_a, \, y_b]$. Then, $k_a$ and $k_b$ are computed using Eq. (\ref{kline}),
\begin{equation}
k_a = \left\lfloor\dfrac{y_a - q}{m}\right\rfloor + 1, \qquad \text{and} \qquad k_b = \left\lceil\dfrac{y_b-q}{m}\right\rceil,
\end{equation}
where $\lfloor x \rfloor$ is the greatest integer lower than $x$ and $\lceil x \rceil$ is the lowest integer greater than $x$. The indexes $k_a$ and $k_b$ are now used to find the elements of $\B{y}$ that are in the interval $[y_a, \: y_b]$. Let $\{k_a:k_b\}$ be the set of integer indexes from $k_a$ to $k_b$, that is, $\{k_a:k_b\} = \{k_a,k_a+1,\cdots,k_b\}$. Then, the elements of $\B{y}$ that are in the interval $[y_a, \: y_b]$ are,
\begin{equation}
    \{\B{y}(i) \in \B{y}\mid \B{y}(i) \in [y_a, \: y_b]\} = \B{y}(\B{I}(k_a:k_b)).
\end{equation}

Using this procedure it may be possible that some extraneous elements -- the closest to the searching range -- would be included in the data retrieved. Since there are $(n - 1)$ number of $\B{y}_l(i)$ bins/steps and $n$ number of elements, there are an average of $E_0 = n/(n - 1)$ elements in each $\B{y}_l(i)$ bin. This means that the expectation for the number of these extraneous element is $n/(n - 1)$, a value that is close to one for large databases. This happens because the two external bins each have a $50\%$ probability that they will contain elements lower than $y_a$ or higher than $y_b$. In case these elements (the closest to the $[y_a, y_b]$ range) cannot be tolerated (strict bounds), they can be removed from the retrieved data in two local searches,
\begin{equation}\label{out}
    \B{y}(\B{I}(k_a \rightarrow)) < y_a \qquad \text{and} \qquad \B{y}(\B{I}(\leftarrow k_b)) > y_b,
\end{equation}
by increasing the indexes from $k_a$ and decreasing the indexes from $k_b$ as long as the inequalities in Eq. (\ref{out}) are satisfied. Thus, other than removing an average of $n/(n - 1)$ elements, this technique does not require performing any kind of search. This fact highlights the most important property of the \kv: \emph{the algorithm complexity is not a function of the database size}.

\subsection{Function inversion using \kv}\label{sec:roots}

This section shows the procedure to obtain the roots of a given nonlinear function $y = f(x)$ using the \kv\ technique within the ranges $[x_{\min}, \, x_{\max}]$ and $[y_{\min}, \, y_{\max}]$. This methodology requires a preprocessing effort where the nonlinear function is discretized to create a database that is first sorted and then accessed using the \kv. This preprocessing effort is done only once for each nonlinear function considered. Therefore, the proposed method is suitable to built a toolbox capable of inverting any nonlinear function, such as Airy, Bessel, beta or Jacobi elliptic functions, using the function evaluation code, $y = f(x)$. On the other hand, this method is not suitable when the function to invert changes continuously.

First, the process requires performing a function discretization in $x$ within the domain of interest using a uniform distribution. Let $n$ be the number of elements of the database and let $\B{x}$ and $\B{y}$ be the vectors containing the database elements, where,
\begin{equation}
    \B{y}(i) = f(\B{x}(i)), \quad \text{with} \quad i = \{1,2,\cdots,n\}.
\end{equation}
Second, the maximum absolute difference, $\delta$, between two consecutive values of $\B{y}$ is computed,
\begin{equation}
    \delta = \max\left(\left|\B{y}(i + 1) - \B{y}(i)\right|\right) + 4 \varepsilon \quad \text{with} \quad i \in [1,n-1].
\end{equation}
This parameter is used to set the \kv\ searching range so that at least one discrete point is found close to each root. Let $y_r$ be the value of the function to be inverted. Then, the function whose roots are required to be computed is,
\begin{equation}
    f(\{x_r\}) - y_r=0,
\end{equation}
where the set of roots for the value $y_r$ is denoted by $\{x_r\}$. Thus, and in order to always retrieve at least one point near each root, the minimum searching range must be,
\begin{equation}
    [y_a, \; y_b] = \left[y_r - \dfrac{\delta}{2}, \; y_r + \dfrac{\delta}{2}\right].
\end{equation}
Third, the $\B{y}$ table is sorted in ascending mode and the \kv\ is built based on this sorted table. Using the searching range $[y_a, \; y_b]$, the \kv\ retrieves the two indexes $k_a$ and $k_b$, that are used to generate the vector $\B{k} = \{k_a:k_b\}$, which contains all the indexes that are required to be retrieved from the database. From vector $\B{k}$ it is possible to obtain the elements of the database in the searching range, in particular, they correspond to the set $\{\B{x}(\B{I}(\B{k}))\}$ and $\{\B{y}(\B{I}(\B{k}))\}$, where $\B{I}$ is the index sorting vector. These elements are all close to the roots, however, they are not in order.

At this point there is no information on the number of roots nor their locations. To solve this, the set $\{\B{x}(\B{I}(\B{k}))\}$, whose size is generally very small\footnote{Using the optimal \kv\ the size of $\{\B{x}(\B{I}(\B{k}))\}$ can be reduced to the number of roots.}, has to be sorted,
\begin{equation}
    \B{x}_s = \{\B{x}(\B{I}_x (\B{I}(\B{k})))\}.
\end{equation}
where $\B{I}_x$ is the sorting index vector between $\{\B{x}(\B{I}(\B{k}))\}$ and $\B{x}_s$. The values of the function in these sorted points ($\B{y}_s$) are,
\begin{equation}
    \B{y}_s = \{\B{y}(\B{I}_x (\B{I}(\B{k})))\}.
\end{equation}

Since the initial discretization is assumed to be uniform, the distance in the variable $x$ between two consecutive points ($\delta_x$) is,
\begin{equation}
    \delta_x = \dfrac{x_{\max} - x_{\min}}{n - 1}.
\end{equation}
Therefore, a fast search can be performed in $\B{x}_s$ to check the number of different roots. If the distance between two consecutive elements in $\B{x}_s$ is smaller than $1.5 \, \delta_x$, that is,
\begin{equation}
    \dfrac{\B{x}_s (i+1) - \B{x}_s (i)}{\delta_x} < 1.5,
\end{equation}
these two points belong to the same root. Conversely, if the distance is bigger, that is,
\begin{equation}
    \dfrac{\B{x}_s (i+1) - \B{x}_s (i)}{\delta_x} > 1.5,
\end{equation}
the two points belong to two consecutive roots. This provides an easy way to count the number of roots, and it also provides a simple criteria to group all the (nearest) points belonging to each root. Then, a root solver such as Newton-Raphson or Regula Falsi can be applied using the closest points retrieved as initial values for the iteration.

\section{Sampling by inversion of the cumulative distribution} \label{sec:cumulative}

The methodology introduced in this section is based on the inverse transform sampling property. Inverse transform sampling consists of generating sample numbers following a probability density function $g (x)$ with $x \in [x_{\min}, x_{\max}]$, given a different probability density function $p (r)$ with $r \in [r_{\min}, r_{\max}]$. Using the definition of probability density functions, we can write,
\begin{equation}
    \ds\int_{r_{\min}}^{r_{\max}} p(\xi) \dd \xi = \ds\int_{x_{\min}}^{x_{\max}} g (\xi) \dd \xi = 1,
\end{equation}
where we are assuming that all the probability is distributed in the defined domain for each  probability density function. Furthermore, a relation between $r$ and $x$ can be obtained through,
\begin{equation}\label{eq01}
    \ds\int_{r_{\min}}^r p(\xi) \dd \xi = \ds\int_{x_{\min}}^x g (\xi) \dd \xi,
\end{equation}
from which it is possible to compute $x = f^{-1} (r)$, where $f$ is the function that relates both variables. This relation can be obtained provided that the two integrals in Eq. (\ref{eq01}) can be computed, and $x$ can be explicitly obtained from the resulting expression of the second integral.

As the most common random number generation algorithms follow a uniform distribution in $[0,1]$, from now on, we suppose that,
\begin{equation}
    \ds\int_{r_{\min}}^r p(\xi) \dd \xi  = U[0,1],
\end{equation}
where $U[0,1]$ is this uniform distribution. Thus, in order to obtain the distribution of points $g (x)$, the inverse of the cumulative distribution of $g (x)$ must be performed, that is, the one that relates variables $x$ and $r$,
\begin{equation}
    U[0,1]= F(\mathbf{x}) = \int_{x_{\min}}^{x} g (\xi) \dd \xi,
\end{equation}
and thus,
\begin{equation}
    \B{x} = F^{-1}(U[0,1]),
\end{equation}
which means that we can obtain samples following the probability density function $g$ by performing a series of inversions of the cumulative distribution of $g$ over a uniform distribution $U[0,1]$.

In order to do these inversions, the \kv\ methodology is used. In particular, we introduce the concept of optimal \kv\ for function inversion to increase the performance of the process by always retrieving the same number of elements from the generated database. This is performed by redistributing the points that constitute the database and that will define the function for the optimal \kv.

\subsection{Optimal \kv\ for function inversion}\label{subsec:fastsearch}

The optimal \kv\ is a modified version of \kv\ that is optimized for function inversion. It is based on the idea of generating a distribution of points such that the number of elements retrieved for a given searching interval is always the same. In order to obtain this feature, an additional preprocessing step that generates the new point distribution is required. However, once the preprocessing is done, the function inversion becomes extremely fast, as the number of points retrieved per root is optimal for the root finder selected (e.g., one for Newton, two for bisection or regula-falsi). The preprocessing for the optimal \kv\ requires three steps.

First, the \kv\ is built using a uniform distribution of points, which is then used in the second step to calculate the roots by following the methodology previously described for function inversion.

Second, the new distribution of points is computed. Let $n_d$ be the size of the optimal \kv\ that we want to define. The parameter $n_d$ can be chosen freely taking into account that; as the value of $n_d$ increases, the memory requirement increases; and as $n_d$ increases, the speed performance of the root solver increases since the points retrieved are closer to the roots. This means that \emph{this methodology can be easily adapted to the requirements of a particular application in terms of performance and memory available}. In addition, $n_d$ represents the number of levels defined in the range of the function. In other words, let $\B{y}_d$ be a uniform distribution over the range of the function, that is,
\begin{equation}\label{y_dis}
    \B{y}_d (i) = y_{\min} + \frac{y_{\max} - y_{\min}}{n_d-1}(i-1),
\end{equation}
where $i = {1, 2,\cdots, n_d}$. Now, we compute all the roots for each value of $\B{y}_d(i)$ using the \kv\ methodology for function inversion. Let $\{\B{x}_d(i)\}$ be the set of roots for each value of $\B{y}_d(i)$. Then, the new distribution of points is given by the set consisting of all the computed roots, that is,
\begin{equation}
    \B{x} = \left\{\bigcup_{i=1}^{n_d}\{\B{x}_d(i)\}\right\}.
\end{equation}
The distribution $\B{x}$ (along with its images $\B{y}=f(\B{x})$) substitutes the original database that is stored in memory.

Third, the \kv\ is built for the new database, $\B{x}$. However, this new \kv\ has an important feature: for an assigned searching range $[y_a, y_b]$, \emph{the number of elements retrieved is always the same}, in particular, let $y_r$ be the value of the function whose roots are required to be computed. Then, the searching range is defined as,
\begin{equation}
\begin{aligned}
    [y_a, \; y_b] &= \left[y_r - n_e\dfrac{\delta}{2}, \; y_r + n_e \dfrac{\delta}{2}\right] \quad \text{where} \quad \\ \delta &= \frac{y_{\max} - y_{\min}}{n_d - 1} + 4 \varepsilon,
\end{aligned}
\end{equation}
and $n_e$ is the number of elements associated with each root. Thus, if $n_e = 1$ only one element per root is obtained, which corresponds to the closest point of the database to each root. On the other hand, if $n_e = 2$ two elements are always retrieved per root, which are the closest elements of the database to the roots and, in addition, these elements bracket the roots. This has two important implications. First, no additional search is required in order to find the closest points from the database to the roots, and second, the points retrieved from the database constitute a very good boundary for applying root solvers such as Newton-Raphson or regula-falsi.

\subsection{Optimal \kv\ applied to random number generation}

For the case of random number generation, the methodology can be improved even further when the number of elements in the database is large enough. In these cases, the process of generating samples can be performed without evaluating the function, and thus, it becomes extremely fast, as the methodology does not require evaluating a function nor performing a search.

Let $y = F(x)$ be the cumulative distribution of the function that the generated samples are required to follow. First, a random number $u$ is computed from a uniform distribution $U[0,1]$ of points, which can be obtained with a random number generation algorithm such as MRG32k3a \cite{ecuyer}. Then, $u$ is transformed into the range $[y_{\min},y_{\max}]$ that the function presents, that is,
\begin{equation}
    y_r = y_{\min} + (y_{\max} - y_{\min}) \, u;
\end{equation}
where $y_r$ is the value whose inverse needs to be computed. Using the optimal \kv, a series of roots are obtained imposing $n_e = 2$ (two points per root). However, it is interesting to note that, as the function studied is a cumulative distribution, it is a monotonically increasing function, and thus, in general, there will be one root per $y_r$ and only two points will be retrieved from the database. This means that the two points obtained are,
\begin{equation}
\begin{aligned}
    (x_b, y_b) &= (\B{x}(\B{k}(1)),\B{y}(\B{k}(1))) \qquad \text{and} \quad \\ (x_t,y_t) &= (\B{x}(\B{k}(2)),\B{y}(\B{k}(2))),
\end{aligned}
\end{equation}
where $\B{k}$ is the vector computed using the optimal \kv\ and that contains the indexes of the database that are in the interval studied. That way, the sample generated can be approximated by a linear interpolation between the two points (as the distances between the elements of the database are small enough) leading to,
\begin{equation}
    x_r = x_b + \ds\frac{y_r - y_b}{y_t - y_b}(x_t - x_b);
\end{equation}
where $x_r$ is the sample generated following the given distribution function.

The process of generating a new sample only requires creating a new random number $u$ and repeating the above process. As it can be seen, following the optimal \kv\ methodology, the function does not need to be evaluated and no real search has to be performed. This means that once the optimal \kv\ is computed and its database is generated, there is no need to store additional information about the function.

However, there are some extreme cases that can appear when the cumulative function has sections of zero derivative, that is, when the distribution function presents intervals of value equal to zero. In order to deal with these cases, the regions with zero derivative must be removed from the random generation, as they do not contribute to the cumulative distribution. Except for this, the methodology is exactly the same as before.

Finally, it is important to note that the performance of this random generation methodology is given by the number of points in the database, $n_d$. If more memory is available (larger $n_d$), then the points retrieved in the inversion process are closer to the roots, and therefore, the inverses are closer to the real values of the function. Another possibility that provides further accuracy is the use of a Newton-Raphson method, which can use the element obtained from the optimal \kv\ when imposing $n_e = 1$ as the starting point for the iteration. However, this methodology requires more computation, as the function is evaluated.

\subsubsection{Example application}

One example application is sampling over a normal distribution. Let $y = f(x)$ be the Gaussian integral defined by,
\begin{equation}
    y = \int_{-\infty}^{x}\frac{1}{\sigma\sqrt{2\pi}}e^{-\frac{\left(t-\mu\right)^2}{2\sigma^2}} \, \dd t,
\end{equation}
where $\sigma$ is the standard deviation and $\mu$ is the mean value of the distribution, and where the function represents the cumulative distribution of the probability density function. This integral can also be expressed by means of the error function $\text{erf}(x)$,
\begin{equation}
\begin{aligned}
    y &= \dfrac{1}{2} \left[1 + \text{erf}\left(\frac{x-\mu}{\sigma\sqrt{2}} \right)\right] \qquad \text{where} \qquad \\ \text{erf}(z) &=  \frac{2}{\sqrt{\pi}}\int_0^z e^{-t^2} \, \dd t,
\end{aligned}
\end{equation}
leading to the expression,
\begin{equation}\label{int_gauss}
    y = \frac{1}{2} \left[1 + \frac{2}{\sqrt{\pi}} \int_0^{\frac{x - \mu}{\sigma\sqrt{2}}} e^{-t^2} \, \dd t\right],
\end{equation}
which is the one that is used in this example to evaluate the Gaussian integral. Equation~\eqref{int_gauss} is defined in all $\mathbb{R}$, however, for practical purposes, the interval $x \in [-5\sigma, \, +5\sigma]$, where $\sigma = 0.2$ and $\mu = 0$, is selected.

\begin{figure}[!h]
	\centering\includegraphics[width=\linewidth]{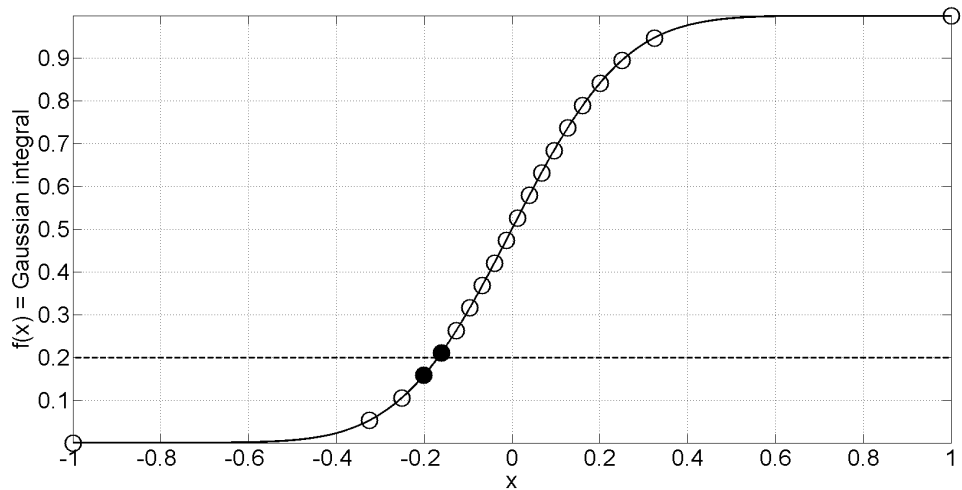}
	\caption{Optimal \kv\ search for the Gaussian integral.}
	\label{fig:gauss1}
\end{figure}

Now, we define the levels of the function using Eq.~\eqref{y_dis} and compute the roots related to them using the general methodology for function inversion. The roots obtained are the new distribution of points and are used to build the optimal \kv. Fig.~\ref{fig:gauss1} shows the new distribution for this example when $20$ levels are defined ($n_d = 20$), where the circles represent the points of the distribution used in the optimal \kv.

Once the preprocessing is done, the calculation of the roots can be made using the general procedure. As an example, let $y_r = 0.2$ be the value of the function whose roots are required to be computed. By using the methodology presented, a value of $x_r=-0.16832$ is obtained. Figure~\ref{fig:gauss1} shows the distribution of points (circles) and the elements retrieved using the \kv\ (filled circles). As it can be seen, the number of points retrieved was two, which was the property that was sought in the development of the optimal \kv\ methodology.

This procedure is continued for each value of $y_r$ until the number of requested points is obtained. As was previously stated, the values of $y_r$ are computed from a uniform distribution $U[0,1]$ of points which can be obtained with a random number generation algorithm such as MRG32k3a \cite{ecuyer}. This means that each time a sample is requested, a function inversion is required. However, the preprocessing required to build the optimal \kv\ is done only once before the start of the process.
\begin{figure}[!h]
	\centering\includegraphics[width=1.0\linewidth]{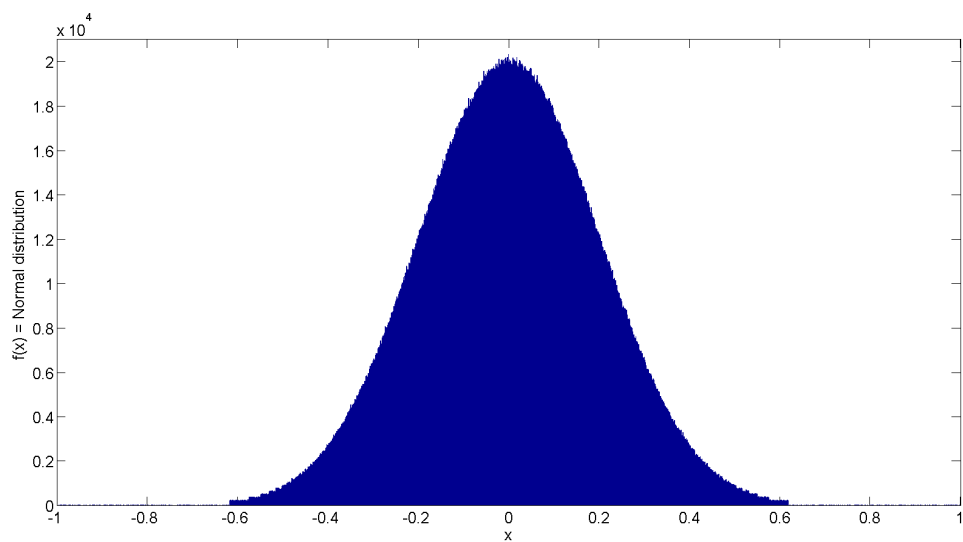}
	\caption{Normal distribution of $10^7$ samples.}
	\label{fig:normal}
\end{figure}

In order to show an application of this method, let the size of the optimal \kv\ be $n_d = 10^3$ and let the probability density function be the normal distribution with standard deviation of $\sigma =0.2$ and mean value of $\mu = 0$ as before. In addition, it is supposed that $10^7$ samples are required to be generated. Figure~\ref{fig:normal} shows the histogram of the samples using the optimal \kv\ methodology. As it can be seen, the samples follow the normal distribution as expected.

Finally, and regarding the speed of the methodology in this example, it took on average 56 seconds to generate the $10^7$ random samples in ten different runs of the program. These numerical tests were performed on a computer with an Intel i7-4600U processor running at 2.1 GHz, 16 GB of RAM, Windows 7 operating system, and  a single threaded script coded in Matlab.

\subsection{Speeding up the algorithm}\label{subsec:speed}

The speed performance of the methodology presented can be increased significantly by doing some modifications. In particular, two improvements are introduced; one consists of a modification in the use of the optimal \kv, while the other is based on a change in the definition of the \kv\ line.

The functions in study are cumulative distributions, therefore, they are monotonic. Thus, for a database created following the optimal \kv\ methodology, it is easy to see that no matter the function in study, the resultant optimal \kv\ has this form: $\B{k}_v = \{0,1,2,3,\dots\}$. This means that the structure of the \kv\ is already known, and thus, it is no longer required to call it during the computation of the samples. However, it is still required to generate the database and the \kv\ line as explained before.

On the other hand, the speed performance of the searching process can be improved even further by defining the \kv\ line as its inverse, in particular:
\begin{equation}
    \begin{aligned}
    m &= \dfrac{n - 1}{y_{\max} - y_{\min} + 2\delta\varepsilon}, \quad \\
    q &= 1 - \dfrac{n - 1}{y_{\max} - y_{\min} + 2\delta\varepsilon}(y_{\min} - \delta\varepsilon).
    \end{aligned}
\end{equation}
Therefore, the elements of the database that bracket the roots can be defined by:
\begin{equation}
    \begin{aligned}
    (x_b, y_b) &= (\B{x} (b), \B{y} (b)) \quad \text{and} \quad \\
    (x_t, y_t) &= (\B{x} (t), \B{y} (t)),
    \end{aligned}
\end{equation}
where:
\begin{equation}
    b = \left\lfloor my_r + q\right\rfloor \quad \text{and} \quad t = b + 1.
\end{equation}

As it can be seen, there is no longer a need to access the \kv, and a lower number of computations are required to retrieve the closest elements from the database. This improves the speed performance of the methodology; the next example provides evidence for this claim.

\subsubsection{Example application}

In order to compare results, we consider the same example as before: the Gaussian Integral. For this case, we also aim to generate $10^7$ samples under the same conditions. When applying this methodology, it took 0.625 seconds to generate the $10^7$ samples, with 0.4 seconds spent to perform the inversions and 0.225 to generate the random numbers using the function \emph{rand} in Matlab. As it can be seen, the speed performance of the methodology has greatly improved (two orders of magnitude).

\subsection{Optimal \kv\ generalization with $n_e$ elements retrieved per sample}

If more accuracy is required, the former methodology can be generalized to any number of elements retrieved per root ($n_e$). This is done by the use of a polynomial interpolation of the points retrieved from the database, which provides a better approximation of the root in terms of accuracy with respect to the linear interpolation. The methodology presented in this section takes advantage of two important properties of cumulative distributions. First, they are right continuous in the domain; and second, they are non-decreasing functions. These properties are used to define the interpolating polynomial in the region near the roots without any problem in the definition.

Let $z=g(x)$ be the distribution function that the samples are required to follow. The cumulative distribution can be defined as,
\begin{equation}
    G^*(x) = xz_ {min} + \int_{x_{min}}^{x}g(t)\dd t,
\end{equation}
where $z_{min}$ is the minimum value of the function and the term $xz_ {min}$ is added in order to obtain a cumulative distribution that is positive in all the domain. A cumulative distribution in the range $[0,1]$ ($G(x)$) can be generated by using the maximum and minimum values of $G^*(x)$ ( $G^*_{max}$ and $G^*_{min}$). Equation \ref{eq:CumDistInRange01} shows how to generate this cumulative distribution.
\begin{equation} \label{eq:CumDistInRange01}
    y = G (x) = \ds\frac{G^* (x) - G^*_{min}}{G^*_{max}},
\end{equation}
The samples are obtained by an inversion of this cumulative distribution using the inverse transform sampling property,
\begin{equation}
    x_r = G^{-1} (U[0, 1]),
\end{equation}
where the random uniform distribution is obtained by a random generation algorithm such as MRG32k3a \cite{ecuyer}.

Now, in order to obtain a value of the root ($x_r$), an approximation is performed using the set of elements retrieved from the database. Let $n_e$ be the number of elements retrieved per root using the optimal \kv. Then, an interpolating polynomial of $n_e-1$ order can be generated using this set of points. This polynomial is defined as a function of $y$ instead of as a function of $x$. This is done in order to perform the inversion of the function directly, without requiring any further inversion. Thus, the root can be approximated using a Lagrange polynomial defined as,
\begin{equation}
    x_r = \sum_{i=1}^{n_e}\left[\B{x}(\B{k}(i))\prod_{\substack{1<m<n_e \\ m \neq i}}\ds\frac{y_r - \B{y}(\B{k}(m))}{\B{y}(\B{k}(i)) - \B{y}(\B{k}(m))}\right],
\end{equation}
where $\B{k}$ is the vector containing the indexes of elements that lie in the searching interval by the application of the optimal \kv.

Finally, it is important to note that the methodology presented in this section can be also applied to the inversion of any kind of monotonic function such as cumulative distributions (as seen in this work) or Kepler's equation (as seen in \cite{inversion}).

\subsubsection{Example application}

\begin{figure}[!h]
	\centering\includegraphics[width=1.0\linewidth]{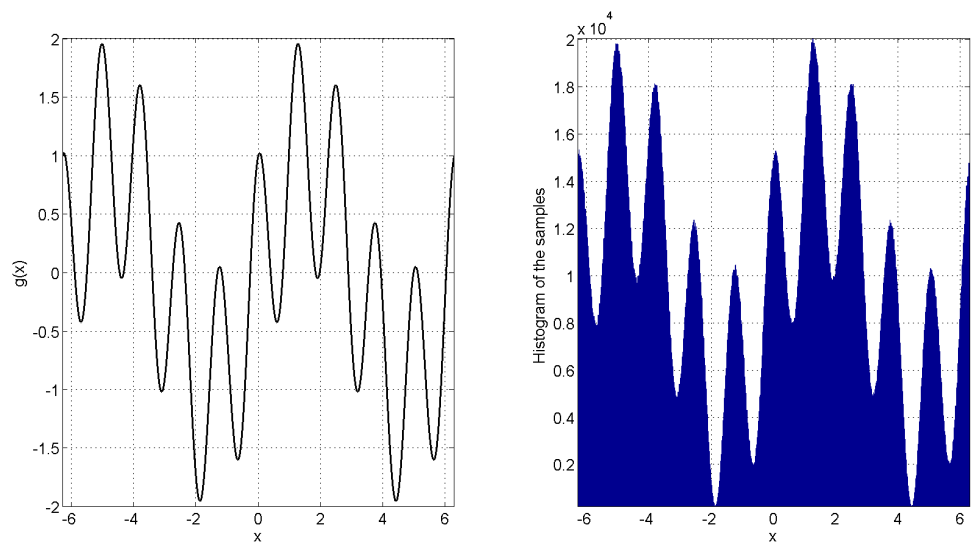}
	\caption{Function and histogram of the distribution of $10^7$ samples.}
	\label{fig:sincos2}
\end{figure}

As an example, let the distribution function be,
\begin{equation}
    g(x) = \sin(x) + \cos(5x),
\end{equation}
where the domain of definition is $x\in[-2\pi,2\pi]$. The representation of this function is shown in Figure~\ref{fig:sincos2} (left). The database and the optimal \kv\ are built using the cumulative distribution of the function and the methodology presented in this section. In addition, and without lost in the generality, we impose $n_e=5$, that is, the number of elements retrieved per root is five. This means that each time that a sample is generated, these five elements are used to generate a Lagrange polynomial of degree four, which is then used to compute the final approximation of the root of the function.

Figure~\ref{fig:sincos2} (right) shows the histogram of $10^7$ samples generated by this procedure. As it can be seen, the histogram follows the distribution function as required. It is important to note that as the parameter $n_e$ increases, the accuracy is increased, but the speed is decreased since more computations are required. For this example, it took 122 seconds to generate all the samples using the same conditions as the former example. Furthermore, when comparing the results using five points ($n_e=5$) with results using two points ($n_e=2$), it was found that it required approximately two times more time with five points than with two points. If instead, we apply the speed improvement shown in Section~\ref{subsec:speed}, and also retrieve five points per root, a time of 48 seconds is required to generate the samples. Finally, if only two points were retrieved using the speed improvement, only 0.650 seconds on average were required to generate all the samples.

This shows three interesting results. First, performing the polynomial interpolation is the process that requires more time to be executed. Secondly, and as shown in the examples, this technique makes it easy to adapt the methodology to the speed and accuracy requirements of each problem. This freedom can be increased due to the possibility of changing the size of both the database and \kv\ used. Thirdly, if using two points and the speed improvement, \emph{the speed performance of the algorithm does not depend on the function selected} (see the comparison between this example and the previous one). This is an expected result from the use of this methodology since the computations required during the inversion are the same no matter the function or the database used.

\section{Sampling by random search in a look up table}

The idea of this alternative methodology is to generate a table containing a large number of points that follows the desired distribution function. Once this preprocessing is done, random retrievals of elements from the table are performed, resulting in a distribution equivalent to the distribution desired.

First, the table that contains the distribution of points has to be built. Let $f(x)$ be the distribution function sought and let $\B{x}$ be a uniform distribution of points between the minimum and maximum values that the samples can take. With this distribution of points, the values of the function $f(x)$ can be easily computed, obtaining $\B{y} = f(\B{x})$, and then the \kv\ can be built using $\B{x}$ and $\B{y}$.

Once the \kv\ is built, a uniform distribution in the variable $\B{y}$ is defined between the minimum and maximum values that the function takes in the interval $x\in [x_{\min},x_{\max}]$, which correspond to the values of $\B{s}(1)$ and $\B{s}(n)$ of the sorted vector, where $n$ is the number of elements of $\B{x}$ and $\B{y}$. That way, the uniform distribution is generated as,
\begin{equation}
    \B{y}_d (i) = \B{s}(1) + \frac{\B{s}(n) - \B{s}(1)}{n_d-1}(i-1),
\end{equation}
where $n_d$ is the number of different values of $y$ considered and $i = \{1,\cdots,n_d\}$ is the integer that defines each one of the values of the function. Then, for each value of $\B{y}_d(i)$, we compute all the roots $x_{r}$ of the distribution function by performing inversions using the \kv\ methodology as seen in Section~\ref{sec:preliminaries} of this work. However, from the set of computed roots, the maximums and minimums obtained are removed, as they are not useful in the distribution process. This is done by checking the two points that bracket the root; If both points are located above (or below) the value $\B{y}_d (i)$, then the root computed is a minimum (or maximum).

The remaining roots along with the maximum and minimum values of $\mathbf{x}$ generate a set of intervals whose boundaries are $[x_{\min}, x_{r1}, x_{r2},\dots, x_{\max}]$. Note that if there are no roots, then there is only one interval defined, $[x_{\min}, x_{\max}]$. Moreover, for the purpose of this problem, we are interested in the intervals where the function lies above the intervals, as we will distribute the points in those intervals. In that respect, we know that when taking two consecutive intervals, the function is positioned above one of them and below the other, because the root defines the point of change of this behavior. Thus, in order to know the position of the function with respect to the intervals, only one check has to be performed corresponding to the first interval, that is,
\begin{eqnarray*}
    \text{if} \ \B{y}(1) & > & \B{y}_d (i) \quad \rightarrow \quad \text{odd intervals selected,} \\
    \text{if} \ \B{y}(1) & < & \B{y}_d (i) \quad \rightarrow \quad \text{even intervals selected.}
\end{eqnarray*}

Now, between the selected intervals, a uniform distribution of points is created in such a way that the distances between points inside an interval are constant and equal to $\Delta x_r$. It is important to note that the value of $\Delta x_r$ must be identical for all the function inversions performed, as this will produce samples that follow the distribution function.

As an example of this procedure, let the boundaries of the intervals be $[0,1,3,4,5]$ for a particular $\B{y}_d (i)$, which means that the probability density function for $\B{y}_d (i)$ has three roots. Then, there exist four different intervals, namely $[0,1]$, $[1,3]$, $[3,4]$ and $[4,5]$. We suppose that $\B{y}(1) < \B{y}_d (i)$, thus, the intervals of interest are $[1,3]$ and $[4,5]$. Now, we create the uniform distribution of points between these intervals, choosing for example $\Delta x_r=0.5$. In this case, the resultant points related to $\B{y}_d (i)$ are $x_{di}=\{1,1.5,2,2.5,3,4,4.5,5\}$. Note that in general, $\Delta x_r$ can be chosen freely or even randomly.

This process in continued for the different values of $\B{y}_d (i)$ and the samples $x_{di}$ stored in memory. Then, each time that a sample is required, a random retrieval is performed using a discrete random number generator such as Alias~\cite{walker} and Vose~\cite{vose}.

The advantage of this methodology is that it is able to generate a large number of samples in a short period of time. However, it has to do an additional retrieval once the database of samples is generated, and thus, this methodology is more suitable when a large population of samples is required to be computed, for example, in Monte Carlo simulations.

\subsection{Example application}

\begin{figure}[!h]
	\centering\includegraphics[width=1.0\linewidth]{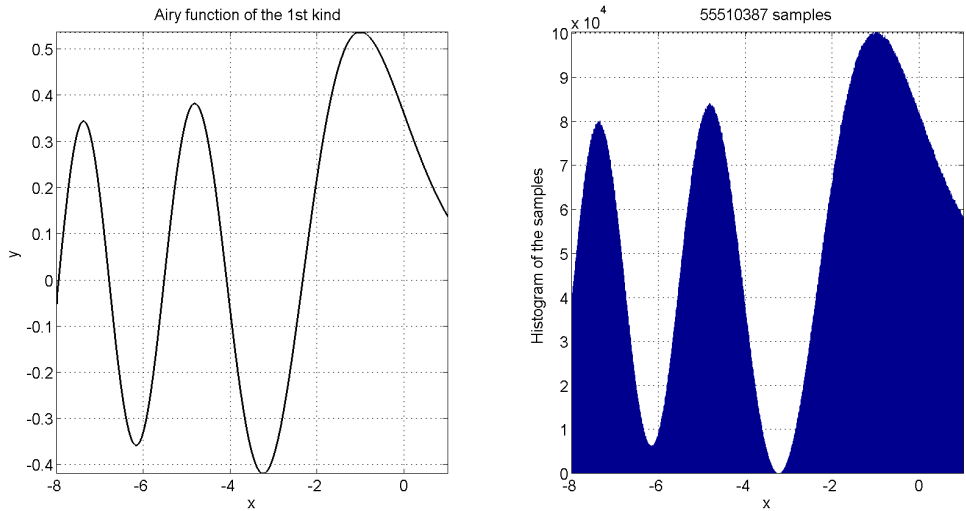}
	\caption{Distribution of samples following an Airy function.}
	\label{fig:airy}
\end{figure}

Let $y=f(x)$ be the Airy function of the first kind, $A_i(x)$, defined as,
\begin{equation}
    y = \dfrac{1}{\pi}\int_{0}^{\infty}\cos\left(\frac{t^3}{3} + xt\right)\, \dd t;
\end{equation}
where it is supposed that a distribution of points is required in the domain $x\in\left[-8,1\right]$. Figure~\ref{fig:airy} (left) shows a plot of the function in this domain of definition.

Now, a large database is generated in order to build the \kv. As the objective is to generate a large number of samples, we select $n = 65,535$ in order to obtain a large \kv\ to perform the inversions quickly. Then, the grid in $x$ and $y$ for the generation of the samples is defined. Let the distance in $x$ between samples in the same interval be,
\begin{equation}
    \Delta x_r = \ds\frac{x_{\max}-x_{\min}}{10,000},
\end{equation}
where it has been supposed that the interval $[-8,1]$ has been divided into $10,000$ elements. Also, we select $n_d = 1,000$ different levels which means that $1,000$ function inversions must be performed during the generation of the database of samples.

Thus, following the methodology explained in this section, a database of with $55,510,387$ elements is generated. The histogram of this distribution can be seen in Fig.~\ref{fig:airy} (right). As it can be observed, the samples follow the distribution function for this example, the Airy function of the first kind. This distribution represents the database from which the samples are now retrieved using a discrete random number generator. Using the same computer and conditions as the other examples, it took an average of 3.4 seconds to generate the $55,510,387$ samples. As it can be seen, both the generation of this database and the preprocessing is faster when compared to the other methodologies, as it requires performing a lower number of function inversions.

Once this preprocessing is performed, the samples are obtained by the retrieval of elements from this database. Alternatively, another possibility is to randomly swap the indexes of the original database, which will lead to the distribution shown in Fig.~\ref{fig:airy}. That way, all the generated samples for the database are used.

\section{Conclusions}\label{sec:conclusions}

The \kv\ is a methodology specially suited for range searching in large databases. In this work, we introduce the application of the \kv\ to probability density function sampling, where the performance properties of the \kv\ can be used to generate a massive number of samples for a given distribution function in a fast and computationally efficient process.

This work presents two different methodologies based on the \kv\ searching technique which are appropriate for massive generation of samples following a fixed distribution function. Both techniques provide a general methodology that can be used to perform transform sampling of any kind of distribution function. Moreover, the methodologies can be adapted to the requirements of memory and CPU power available as the \kv\ size can be appropriately defined to meet those requirements. Possible applications of these techniques include Monte Carlo simulations and the creation of a toolbox for random number generation.

The first of the methodologies presented is based on the inverse transform sampling property, where the performance of the optimal \kv\ can be used for fast function inversion of the cumulative distribution of the distribution function. This methodology requires a preprocessing effort that consists of several function inversions that are used to build the optimal \kv\ and database. However, once this preprocessing is done, the optimal \kv\ is fixed for the desired distribution function, and thus, can be implemented directly for particular applications without requiring the evaluation of the function nor performing searches to generate samples. In addition, and when using the speed improvements introduced in this work, the algorithm allows to obtain samples extremely fast, being its speed performance during the inversion independent of the function considered.

The second methodology allows the fast generation of a large number of samples following a given distribution function. The samples are generated by a series of inversions of the distribution function where points are massively distributed inside the intervals defined by the roots of the function at different values. Then, this distribution is stored and accessed through random searches that retrieve the samples following the original probability density function. The advantage of this technique is that the samples are generated extremely fast. However, the distribution has to be stored in a table for future search, and thus, it requires more memory than the other technique.

\section{Acknowledgments}

The work of D. Arnas was supported by the Spanish Ministry of Economy and Competitiveness (Project no. ESP2017\textendash87113\textendash R).

\bibliographystyle{ieetr}

\end{document}